# TESTABILITY MEASUREMENT MODEL FOR OBJECT ORIENTED DESIGN (TMM<sup>OOD</sup>)

Abdullah[1], Dr. M.H. Khan[2], Dr. Reena Srivastava[3]

[1]Research Scholar, School of Computer Application, BBD University, Lucknow, India
[2]Associate Professor, Department of C.S. E., I.E.T., Lucknow, India
[3]Dean, School of Computer Application, BBD University, Lucknow, India

## ABSTRACT

*Measuring testability early in the development life cycle especially at design phase is a criterion of crucial importance to software designers, developers, quality controllers and practitioners. However, most of the mechanism available for testability measurement may be used in the later phases of development life cycle. Early estimation of testability, absolutely at design phase helps designers to improve their designs before the coding starts. Practitioners regularly advocate that testability should be planned early in design phase. Testability measurement early in design phase is greatly emphasized in this study; hence, considered significant for the delivery of quality software. As a result, it extensively reduces rework during and after implementation, as well as facilitate for design effective test plans, better project and resource planning in a practical manner, with a focus on the design phase. An effort has been put forth in this paper to recognize the key factors contributing in testability measurement at design phase. Additionally, testability measurement model is developed to quantify software testability at design phase. Furthermore, the relationship of Testability with these factors has been tested and justified with the help of statistical measures. The developed model has been validated using experimental tryout. Finally, it incorporates the empirical validation of the testability measurement model as the author's most important contribution.*

## 1. INTRODUCTION

Software industries have realized that quality software is an important means to improve performance and to gain competitive advantage. The development and supply of software is one of the fastest growing industry segments in India and abroad. The software developed as commercial products are generally much different from computer programs written for academic or research purpose. A great deal of effort, time and cost are required to design and develop any commercial software. Development of even a trivial piece of software requires many activities to be performed and it is regarded as a project. Computer programming is just one part of the software development process. Like any other economic endeavor, development of software requires a systematic approach that includes comprehensive methods, better tools for smooth execution of these methods and procedures for quality assurance, coordination and control [1]. Testability has been recognized as one of the most important issues in the field of Software Engineering. It presents insights that are found to be very much important and helpful for the duration of software development life cycle and quality assurance [2] [3] [19].

Testability study early in the development life cycle is a norm of crucial importance to software designers, developers and the quality controllers. However, most of the studies measure testability or precisely the attributes that have impact on testability at the source code level [4] [5]. Measuring testability at a later stage leads to the late arrival of desired information, leading to late decisions about changes in design. This simply increases cost and rework. A decision to change the design in order to improve testability after coding has started may be very expensive and error-prone. Therefore, early evaluation of testability in the development process may enhance quality and reduce testing efforts and costs. Testability estimation early in design phase is highly emphasized in the study; hence, considered important for the delivery of quality software. The major objective of software testability measurement is to find out which software components are inferior in quality, as well as where errors can hide from software testing.

A frame work to measure testability of object oriented software early at the design phase has been proposed, and it has been validated with a sound theoretical basis for high and improved level acceptability [2]. An effort has been put forth to recognize the key factors contributing in testability measurement at design phase of development life cycle. It has been concluded that Modifiability and Flexibility are the two most important factors affecting software testability measurement in design phase. Taking into consideration the importance and significance of their contribution, a testability measurement model has been developed to measure software testability at design phase. Furthermore, statistical test are performed to justify the correlation of testability with Modifiability and Flexibility.

## 2. TESTABILITY FACTORS AND KEY CONTRIBUTORS

Software testability is now established to be an important and distinct software quality characteristic. Testability is one of the most essential quality indicators and its measurement leads to the prospects of facilitating and improving a test process. Testability has always been an elusive concept and its correct measurement or evaluation a difficult exercise [6]. Moreover, there is little consensus among researchers and practitioners about 'what aspects of software are truly related to testability'. So it is hard to get an understandable view on all the prospective factors that have an effect on testability and the dominant degree of these factors under different testing perspectives [8]. Researchers and Practitioners have made significant amount of effort and contribution in the way of investigating testability factors in common and object oriented software in particular [9] [10] [11] [12] [13] [14] [21]. It comes into view fairly, conclusive from the existing literature review that there is a difference among researchers and practitioners in considering the factors while measuring testability in general and absolutely at design phase.
A truthful measure of software quality fully depends on testability measurement, which in turn depends on the factors that have an effect on software testability. Above mentioned explanation confirms that there is a divergence among researchers and practitioners regarding considering the factors while estimating the testability. Hence, it appears extremely desirable and important to recognize the factors that facilitate testability in order to get the accurate and reliable measure of software testability.

Despite the fact that, getting a universally accepted set of testability factors is only probable. Testability quality criteria are the characteristics which help to identify the testability factors. Criteria present a more complete, actual definition of factors as well as criteria common among factors assist to show the interrelationship between factors. The criteria of the testability factor are the characteristics of the software product or development cycle by which the factor can be judged or recognized. An endeavor has been made to collect a set of testability factors that can affect software testability. However, without any loss of generality, it comes into view to include the factors namely, modifiability, simplicity, understandability, flexibility, traceability, complexity, self descriptiveness and modularity as testability factors. Out of these eight factors, some of them have their positive impact in measuring testability of object oriented software design, while others have less or negligible impact. An effort has been made to recognize the testability factors that truly affect testability measurement at design phase. Modifiability and Flexibility are the key testability factors that truly affect software testability measurement and fulfill the quality criteria, particularly Modifiability quality criteria is understandability, traceability, self descriptiveness and Flexibility quality criteria is simplicity, complexity [21]. Therefore, without any loss of generality, it comes into view realistic to include Modifiability and Flexibility for testability measurement at design phase.

## 3. MODEL DEVELOPMENT

The generic quality models [15] [16] have been considered as a basis to develop the Testability Measurement Model for Object Oriented Design (TMM$^{OOD}$) shown in figure 1, which involves the following steps.

I) Recognition of testability factors of object oriented software that has positive impact on testability measurement at design phase of development life cycle.
II) Recognition of object oriented design properties.
III) A means of connecting of them.

Based upon the relationship of the testability factors and design constructs, the relative significance of individual factors that has major impact on testability at design phase is weighted proportionally. In order to set up a model for Testability Measurement, a multiple linear regression method has been used to get the coefficients [17]. This method establishes a relationship between dependent variable and multiple independent variables. Multivariate linear equation is given below in *Eq* (1) which is as follows.

$$Y = a_0 + a_1 X_1 + a_2 X_2 + a_3 X_3 + \text{-------} + a_n X_n \qquad \qquad Eq~(1)$$

Where,
- **Y** is dependent variable.
- X1, X2, X3--------Xn are independent variables, associated to Y and are expected to explain the variance in Y.
- a1, a2, a3--------an., are the coefficient of the particular independent variables.
- And $a_0$ is the intercept.

It has been extensively reviewed and discussed in section 2 that Modifiability and Flexibility are the major factors affecting software testability measurement at design phase. Therefore, these key testability factors were addressed well in advance while integrating testability at design phase. By applying the regression method, we developed Modifiability Model [6] and Flexibility Model [7] that are given below in equation (2) and (3) respectively. The model of Modifiability and Flexibility forms the strong basis for development of testability measurement model.

**Modifiability = 1.107- .102 × Encapsulation + 1.810 × Inheritance + .850 × Coupling**
*Eq* **(2)**

**Flexibility= 1.051 + 2.320 × Encapsulation + 0.160 × Coupling - 2.283 × Cohesion + 11.572 × Inheritance**      *Eq* **(3)**

Metrics of the design constructs namely Encapsulation (ENM), Inheritance (INM), Coupling (CPM) and Cohesion (COM) are used to address the major testability factors namely Modifiability and Flexibility. These two factors are further used to measure and control testability of object oriented software at design phase. Below is the Figure 1, which gives an overview of the main idea.

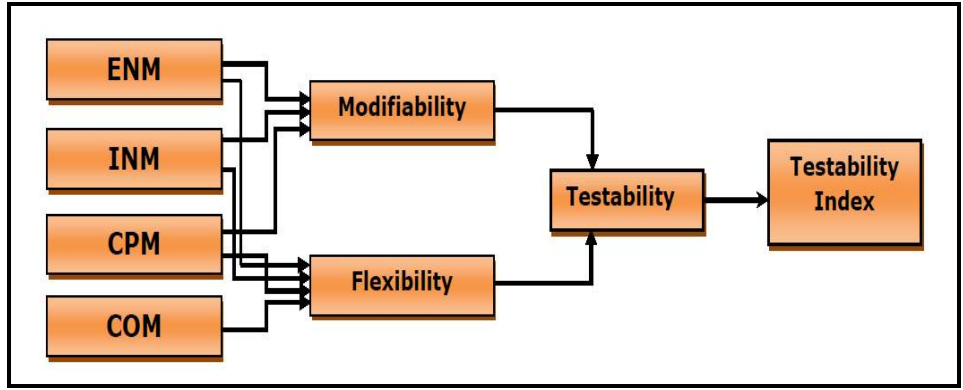

Figure 1: Mapping Design Constructs with Key Factors of Testability

In order to establish a model for testability measurement, a multiple linear regression technique discussed above in equation (1) has been used. Taking into account the impact of Encapsulation, Inheritance, Coupling and Cohesion on testability contributors 'Modifiability and Flexibility', following MR equation has been formulated that can measure testability of object oriented design.

**Testability = $\alpha_0$ + ß$_1$ × Modifiability + ß2 × Flexibility**     *Eq (4)*

The data used for developing model given in equation (4) has been taken from [20], which consist of six industrial projects with around 10 to 20 classes. The values of 'Encapsulation Metrics (ENM), Inheritance Metrics (INM), Coupling Metrics (CPM) and Cohesion Metrics (COM)'and the values of 'Modifiability and Flexibility' have been used. Using SPSS, correlation coefficients are computed and model of testability measurement is thus formulated as given below in equation (5).

**Testability = -98.666 + 49.210 × Modifiability - 2.983 × Flexibility**   *Eq (5)*

Table 1:  Correlation coefficients

| Model | | Unstandardized Coefficients | | Standardized Coefficients | t | Sig. |
|---|---|---|---|---|---|---|
| | | B | Std. Error | Beta | | |
| 1 | (Constant) | -98.666 | 25.518 | | -3.866 | .031 |
| | Modifiability | 49.210 | 11.538 | 1.331 | 4.265 | .024 |
| | Flexibility | -2.983 | 1.768 | -.527 | -1.687 | .190 |

The Model Summary table of the output is most useful when performing multiple regression. Capital R is the multiple correlation coefficients that tell us how strongly the multiple independent variables are related to the dependent variable. R square is very supportive as it gives us the coefficient of determination. The Model Summary is shown in Table 2.

Table 2:  Model Summary

| Model | R | R Square | Adjusted R Square | Std. Error of the Estimate |
|---|---|---|---|---|
| 1 | .950[a] | .903 | .839 | 5.81531 |

# 4. VALIDATING the MODEL -TMM$^{OOD}$

The applications that are used in validating the multivariate linear regression model for computation of Testability (*Eq.* 5) have been taken from [18].We labelled the applications as: System W, System X, System Y, and System Z. All the systems are commercial software implemented in C++ with the number of projects as shown in table 3(a).

Table 3(a): Group and Projects

| Group | Projects |
|-------|----------|
| System W | 6 |
| System X | 4 |
| System Y | 7 |
| System Z | 4 |

Table 3(b): Descriptive Statistics for System W

|  | Minimum | Maximum | Mean |
|--|---------|---------|------|
| Modifiability | 5.86 | 9.89 | 8.2029 |
| Flexibility | 3.78 | 9.29 | 7.3063 |
| Testability | 171.27 | 360.09 | 283.2014 |

Table 3(c): Correlations Analysis for System W

|  | Testability | Modifiability | Flexibility |
|--|-------------|---------------|-------------|
| Testability | 1 | .999 | .877 |
| Modifiability | .999 | 1 | .893 |
| Flexibility | .877 | .893 | 1 |

Table 3(d): Descriptive Statistics for System X

|  | Minimum | Maximum | Mean |
|--|---------|---------|------|
| Modifiability | 2.64 | 5.85 | 3.9059 |
| Flexibility | 4.78 | 7.58 | 6.4951 |
| Testability | 17.08 | 166.76 | 74.1701 |

Table 3(e): Correlations Analysis for System X

|  | Testability | Modifiability | Flexibility |
|--|-------------|---------------|-------------|
| Testability | 1 | .999 | .772 |
| Modifiability | .999 | 1 | .794 |
| Flexibility | .772 | .794 | 1 |

Table 3(f): Descriptive Statistics for System Y

|  | Minimum | Maximum | Mean |
|---|---|---|---|
| Modifiability | 2.34 | 3.13 | 2.7666 |
| Flexibility | 3.56 | 10.47 | 7.2894 |
| Testability | 4.50 | 29.56 | 15.7328 |

Table 3(g): Correlations Analysis for System Y

|  | Testability | Modifiability | Flexibility |
|---|---|---|---|
| Testability | 1 | .955 | .763 |
| Modifiability | .955 | 1 | .921 |
| Flexibility | .763 | .921 | 1 |

Table 3(h): Descriptive Statistics for System Z

|  | Minimum | Maximum | Mean |
|---|---|---|---|
| Modifiability | 1.61 | 3.46 | 2.9344 |
| Flexibility | 1.41 | 8.30 | 5.9949 |
| Testability | -23.53 | 48.24 | 27.8534 |

Table 3(i): Correlations Analysis for System Z

|  | Testability | Modifiability | Flexibility |
|---|---|---|---|
| Testability | 1 | .999 | .987 |
| Modifiability | .999 | 1 | .992 |
| Flexibility | .987 | .992 | 1 |

Table 3(j): Correlations Analysis Summary

|  | Testability × Modifiability | Testability × Flexibility |
|---|---|---|
| System W | .999 | .887 |
| System X | .999 | .772 |
| System Y | .955 | .763 |
| System Z | .999 | .987 |

Table 3 (j) summarizes the result of the correlation analysis for testability measurement model, which shows that for all the system, both Modifiability and Flexibility are highly correlated with testability. The value of correlation 'r' lies between ±1, positive value of 'r' in table 3(j): designates positive correlation between the two variables. The values of 'r' close to +1 specify high degree of correlation between the two variables in above table 3(j).

## 5. EMPIRICAL VALIDATION

The empirical validation is an essential stage of planned research. Empirical validation is the correct approach and practice to justify the model acceptance. Keeping view of this fact, realistic validation of the testability measurement model has been performed using sample tryouts. In order to validate developed testability measurement model the data has been taken from [18]. During experiments, testability value of the projects has been calculated using the developed model, followed by the calculation of testability rating. These calculated ratings are then compared with the known rating given by experts with the help of Charles Speraman's Coefficient of Correlation.

Table 4: Computed Ranking, Actual Ranking and their Relation

| Projects ↓ | Testability Value | | Testability Ranking | | $\sum d^2$ | $r_s$ | $r_s$ >.4815 |
|---|---|---|---|---|---|---|---|
| | Computed Value | Known Value | Computed Ranking | Known Ranking | | | |
| p1 | 199.919 | 5.879 | 19 | 19 | 0 | 1.0000 | ✓ |
| p2 | 280.913 | 7.716 | 20 | 20 | 0 | 1.0000 | ✓ |
| p3 | 343.761 | 9.188 | 22 | 22 | 0 | 1.0000 | ✓ |
| p4 | 360.092 | 9.565 | 23 | 23 | 0 | 1.0000 | ✓ |
| p5 | 343.249 | 9.174 | 21 | 21 | 0 | 1.0000 | ✓ |
| p6 | 48.877 | 2.268 | 15 | 15 | 0 | 1.0000 | ✓ |
| p7 | 63.969 | 2.615 | 16 | 16 | 0 | 1.0000 | ✓ |
| p8 | 166.758 | 5.013 | 17 | 17 | 0 | 1.0000 | ✓ |
| p9 | 171.274 | 5.022 | 18 | 18 | 0 | 1.0000 | ✓ |
| p10 | 17.076 | 1.599 | 7 | 10 | 9 | 0.9956 | ✓ |
| p11 | 4.498 | 1.176 | 3 | 3 | 0 | 1.0000 | ✓ |
| p12 | -9.151 | 0.832 | 2 | 2 | 0 | 1.0000 | ✓ |
| p13 | 18.055 | 1.490 | 9 | 6 | 9 | 0.9956 | ✓ |
| p14 | 9.692 | 1.294 | 5 | 5 | 0 | 1.0000 | ✓ |
| p15 | 29.560 | 1.772 | 11 | 11 | 0 | 1.0000 | ✓ |
| p16 | 14.900 | 1.532 | 6 | 7 | 1 | 0.9995 | ✓ |
| p17 | 48.235 | 2.242 | 14 | 13 | 1 | 0.9995 | ✓ |
| p18 | 46.873 | 2.242 | 13 | 14 | 1 | 0.9995 | ✓ |
| p19 | 19.541 | 1.577 | 10 | 9 | 1 | 0.9995 | ✓ |
| p20 | 17.582 | 1.547 | 8 | 8 | 0 | 1.0000 | ✓ |
| p21 | 5.993 | 1.243 | 4 | 4 | 0 | 1.0000 | ✓ |
| p22 | 39.839 | 2.041 | 12 | 12 | 0 | 1.0000 | ✓ |
| p23 | -23.533 | 0.600 | 1 | 1 | 0 | 1.0000 | ✓ |

This table (4), therefore, indicates a very significant correlation between the computed ranking and actual ranking of testability model, at the 0.01 for a 99% confidence interval.

➔ $r_s$ >0.4815 means significant results.
➔ Testability measurement model had statistically significant correlations with 23 of 23 projects.

As mentioned above, Charles Speraman's Coefficient of Correlation (rank relation) $r_s$ was used to check the significance of correlation between 'Calculated Values' of Testability using model and it's 'Known Values'. Rank correlation is the process of determining the degree of correlation between two variables. The '$r_s$' was calculated using the method given as under: Speraman's Coefficient of Correlation

$$r_s = 1 - \frac{6\sum d^2}{n(n^2 - 1)} \qquad -1.0 \le r_s \le +1.0$$

'd' = difference between 'Calculated ranking' and 'Known ranking' of testability.
 n = number of projects (n=28) used in the experiment.

The correlation values between testability through model and known ranking are shown in table (4) above. Pairs of these values with correlation values $r_s$ above [±.4815] are checked in critical values table. The correlations are up to standard with high degree of confidence, i.e. up to 99%. Therefore we can conclude without any loss of generality that testability measurement model; measures are really highly reliable and significant at design phase.

## 6. HYPOTHESIS TESTING OF COEFFICIENT OF CORRELATION

An experimental coefficient of corelation of Modifiability and Flexibility with Testability strongly indicates the higher importance and significance of taking into consideration both the key factors (Modifiability and Flexibility) for making a measurement of software testability at design phase. Furthermore, to justify the claim, a test to conclude the statistical significance of the correlation coefficient observed possibly will be appropriate. For the motivation, Hypothesis testing is performed to test the significance of r (Correlation Coefficient) using the given below formula:

$$T = \frac{r\sqrt{N-2}}{\sqrt{1-r^2}}$$

With N-2 degree of freedom, a coefficient of correlation is evaluated as statistically importance when the t value equals or exceeds the t critical value in the t distribution critical values.

$H_{0\,(T^{\wedge}M)}$: **Testability and Modifiability are not highly correlated.**

Table 5(a): Correlation Coefficient Test for Testability and Modifiability

|  | **System W** | **System X** | **System Y** | **System Z** |
|---|---|---|---|---|
| Testability ^ Modifiability | .999 | .999 | .955 | .999 |
| $t_r$ | 44.69 | 31.60 | 7.20 | 31.60 |
| $t_{r\text{-Critical Value}}$ | 2.447 | 2.776 | 2.365 | 2.776 |
| $t_r > t_{r\text{-Critical Value}}$ | ✓ | ✓ | ✓ | ✓ |
| $H_0(T^\wedge M)$ | Reject | Reject | Reject | Reject |

$H_{0\,(T^\wedge F)}$: **Testability and Flexibility are not highly correlated.**

Table 5(b): Correlation Coefficient Test for Testability and Flexibility

|  | **System W** | **System X** | **System Y** | **System Z** |
|---|---|---|---|---|
| Testability ^ Flexibility | .877 | .772 | .763 | .987 |
| $t_r$ | 3.65 | 1.72 | 2.64 | 8.68 |
| $t_{r\text{-Critical Value}}$ | 2.447 | 2.776 | 2.365 | 2.776 |
| $t_r > t_{r\text{-Critical Value}}$ | ✓ | × | ✓ | ✓ |
| $H_0(T^\wedge F)$ | Reject | Accept | Reject | Reject |

Using 2-tailed test at the 0.05 for a 95% confidence interval with different degrees of freedom, it is clear from the tables 5(a) and (b), the null hypothesis is rejected (with the exception of, for system 'X' of Testability and Flexibility).As a result, the researcher's claim of correlating Testability with Modifiability and Flexibility at design phase is Statistically extremely justified.

# 7. CONCLUSIONS

In this paper, software testability key factors are identified and their impact on testability measurement and improvement at design phase has been analyzed. 'Modifiability and Flexibility', two of the key factors affecting object oriented design testability have been taken into consideration. Considering both the major factors, a testability measurement model for object oriented design has been developed (TMM$^{OOD}$), and the statistical inferences are validated for high level better acceptability. The developed model to measure testability of object oriented software design is extremely trustworthy and correlated with object oriented design constructs. Testability measurement model has been validated theoretically as well as empirically using experimental try-out. The applied validation on the testability model concludes that proposed model is highly consistent, acceptable and reliable.

## ACKNOWLEDGEMENTS

First and foremost, I would like to thank my supervisor Dr. Reena Srivastava and Dr. M.H. Khan, for standing beside me throughout my research and authoring this research paper. Both has been my inspiration and motivation for continuing to improve my knowledge and move my research forward.

## Authors

**Abdullah** received the MCA degree from Uttar Pradesh Technical University, Lucknow, in 2006. He is currently working as an Associate Professor, in the Department of Computer Application, at Institute of Environment and Management, Lucknow. His research interests Include Software testability, Software Quality Estimation. He has written various books and study materials for (North Orissa University) Orissa, (Suresh Gyan Vihar University, Jaipur) Rajasthan, (Bharati Vidyapeeth University, Pune) Maharashtra, NAAC Re-Accredited "A" Grade University.

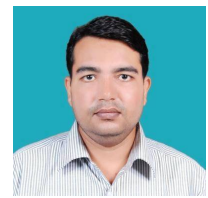

**Dr. M. H. Khan**, Associate Professor, Department of Computer Science and Engineering at IET Lucknow UP. He obtained his MCA degree from Aligarh Muslim University (A Central University) in 1989. Later he did his PhD from Lucknow University. He has around 25 years rich teaching experience at UG and PG level. His area of research is Software Engineering. Dr. Khan published numerous articles, several papers in the National and International Journals and conference proceedings.

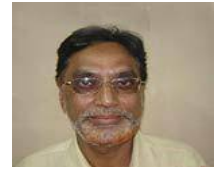

**Dr. Reena Srivastava** is currently working as Dean, School of Computer Applications at BBD University, Lucknow. She received her Ph.D. degree from MNNIT Allahabad, India. Her research area includes Multi-Relational Classification, Privacy Preserving Data Mining and Software Engineering.

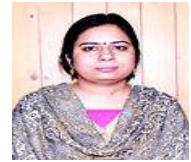